\documentclass[journal]{IEEEtran}
\ifCLASSINFOpdf
\else
\fi
\usepackage{graphicx}
\usepackage{caption}
\graphicspath{{./images/}}
\begin{document}

\title{IoT Security Challenges and Mitigations: An Introduction}

\author{Stuart Millar, Rapid7 LLC, \textit{stuart\_millar@rapid7.com}% <-this % stops a space

\thanks{S. Millar is with Rapid 7 LLC, Boston, MA, USA.  This version dated October 21st 2016.  For correspondence please e-mail: stuart\_millar@rapid7.com}% <-this % stops a space
%\thanks{Manuscript received Month Day, Year tbc; revised Month Day, Year tbc.}
}

\maketitle


\begin{abstract}
The use of IoT in society is perhaps already ubiquitous, with a vast attack surface offering multiple opportunities for malicious actors.  This short paper first presents an introduction to IoT and its security issues, including an overview of IoT layer models and topologies, IoT standardisation efforts and protocols. The focus then moves to IoT vulnerabilities and specific suggestions for mitigations. This work's intended audience are those relatively new to IoT though with existing network-related knowledge.  It is concluded that device resource constraints and a lack of IoT standards are significant issues.  Research opportunities exist to develop efficient IoT IDS and energy-saving cryptography techniques lightweight enough to reasonably deploy.  The need for standardised protocols and channel-based security solutions is clear, underpinned by legislative directives to ensure high standards that prevent cost-cutting on the device manufacturing side.  
\end{abstract}


% \fi
\IEEEpeerreviewmaketitle

\section{Introduction}
% for IEEE journal papers produced under \LaTeX\ using
% IEEEtran.cls version 1.8b and later.

\IEEEPARstart{T}he term ‘Internet Of Things’ (IoT) was created by Kevin Ashton in 1999 and then formally introduced by the International Telecommunication Union in 2005.  The Cluster of European Research Projects defines IoT as allowing people and things to be connected anytime, anyplace, with anything and anyone, ideally using any path/network and any service.  ‘Things’ are actual objects, for example smart wearables, thermostats, or sensors detecting speed and light \cite{shah}.  These IoT devices can access the cloud, transfer data and be controlled remotely, thus acting as a gateway to the internet.  By 2020 it is estimated 4.5 billion new people and 37 billion new objects will be part of the IoT \cite{ciscoiot}.  The term 'Network of Things' (NoT) also exists, and it can be said the overall IoT is made up of various NoT.  

The nature of IoT devices often grant them access to our sensitive personal data, producing a large attack surface with everything from fridges to sprinklers being wired and interconnected.  A recent industry report \cite{mowbrayhp} analysed devices from manufacturers of TVs, webcams, door locks, home alarms and more, revealing some concerning results – 70\% of them used unencrypted network services and the majority failed to encrypt network services transmitting data via the internet and the local network.  Indeed, \cite{mowbrayhp} went on to say users are only one network misconfiguration away from exposing their data to the world via wireless networks.  The widening spread of IoT sensors increases this attack surface at home, in the office, and critically with industrial control systems (ICS).  The Stuxnet malware worm, which caused massive disruption to Iranian nuclear centrifuges, is an example of an IoT compromise.  In the US, the National Cybersecurity and Communications Integration Center (NCCIC) has a specialist ICS team that reports on such incidents \cite{6996632}, highlighting their importance.    
 
% I wish you the best of success.

 

% \textcolor{blue}{NB cite previous from eventual submission venue.} 

Security is tough in resource-constrained environments.  IoT sensor nodes typically have slow processors, limited computational power and little memory storage, often comprising an 8-bit processor, RAM measured in KB rather than MB, a small radio and tiny battery.  IoT networks regularly use low-energy technologies such as IEEE 802.15.4, bluetooth and low-power wi-fi \cite{ndn}, operating with a smaller maximum transmission unit (MTU) and lower transmission rate than normal ethernet links.  There are unique technical challenges in keeping IoT packet sizes small and overheads low.  IoT resource constraints cause reliance on lossy and low-bandwith channels, meaning resource-intensive cryptography functions are of limited value too \cite{rfc8576}.  

The selection of topology has considerable impact. NoT can have points of centralised control called base stations, which are gateways to other networks, data storage/processing centres or access points for a human interface \cite{1203362}.  Traditional networks are point-to-point, whereas NoT traffic can be many-to-one (sending data from nodes to a base), one-to-many (base to nodes communication, like a multicast or a message flood) and local (neighboring nodes talking to each other).  

In practice, IoT deployments are found in exposed, physically insecure or hostile environments open to attack \cite{7575852}, where classic physical security challenges exist - the simple lack of manual human control means the devices need managed and protected \cite{shah}, while vulnerable Radio Frequency Identification (RFID) usage is common \cite{xingmei}.  More complex IoT-specific attacks include the sinkhole attack, the HELLO flood, the wormhole attack and the Sybil attack \cite{1203362}.

\begin{figure*}%[b]
    \centering
    \captionsetup{justification=centering}
    \captionsetup{font=small}
    \includegraphics[width=\linewidth]{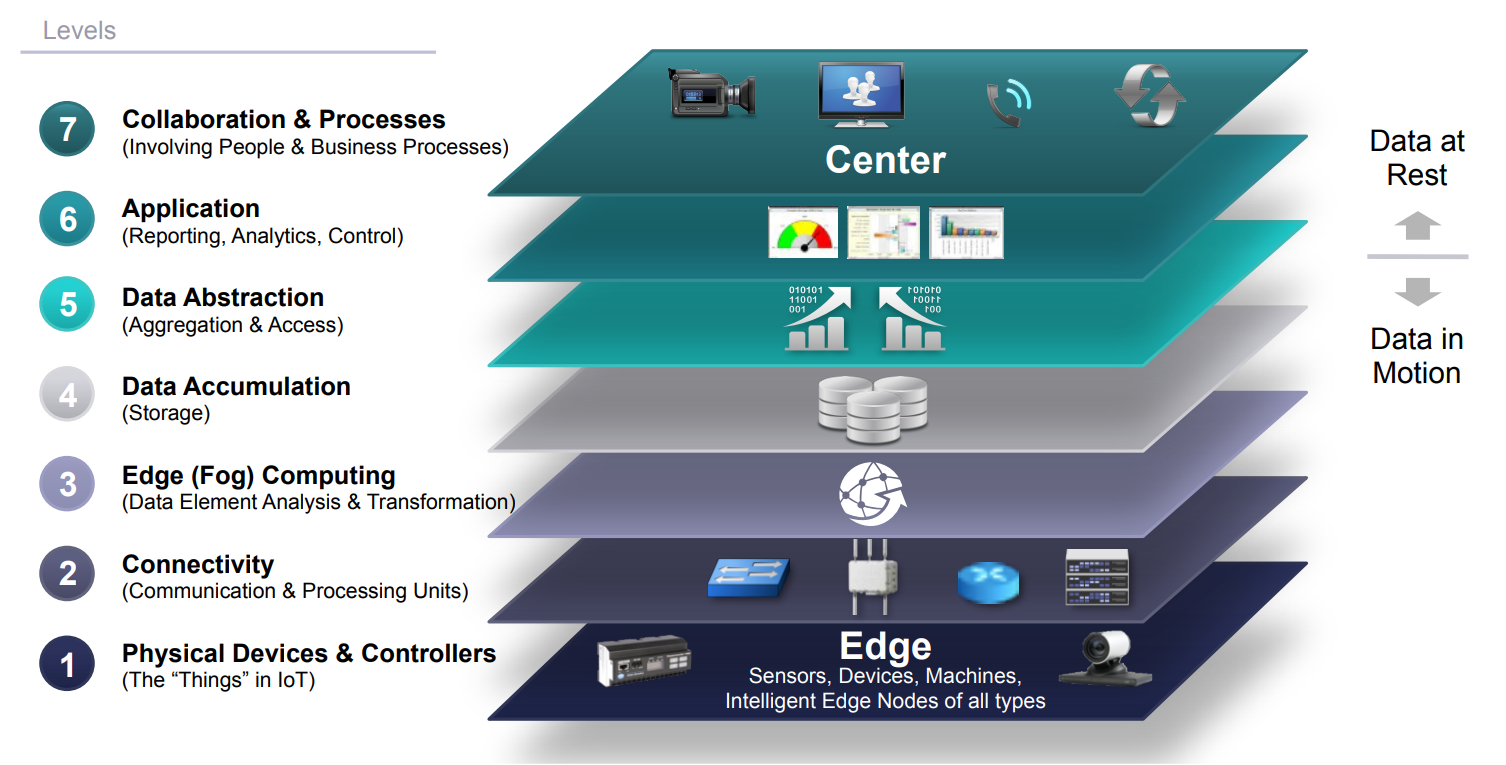}
    % \vspace{-1cm}
         \caption{CISCO's IoT reference model \cite{ciscoiot}}
    \label{fig:ciscoiotmodel}
\end{figure*}

This paper contributes survey-style introductory discussion on:

\begin{itemize}
  \item IoT layer models, topologies, protocols and the lack of standards across the same.
  \item Vulnerabilities that exist in IoT, associated IoT-specific attacks and more conventional attack vectors.
  \item Mitigations and countermeasures that can be deployed.
\end{itemize}

It is organised as follows:  Section II gives an overview of two IoT layer models, Section III outlines the benefit of a mesh topology and Section IV presents various IoT protocols.  Section V contains IoT vulnerabilities and attacks, with Section VI detailing mitigations.  Lastly Section VII offers conclusions and future directions.

%\IEEEpubidadjcol

\section{IoT layer models}
\label{subsec:methodology}

An seminal paper from \cite{1203362} concerned secure routing in IoT, attacks and countermeasures, including comparing IoT with more traditional wireless networks while noting resource constraints in IoT.  At the time of writing, there is no single standard and unified IoT layer model. Different layer models have been proposed \cite{mahmoud, ciscoiot}, in a similar fashion to the well-known OSI model for networking.   

\cite{mahmoud} present a simple model of three layers - perception, network and application:

\begin{enumerate}
    \item \textit{Perception layer} – consider this a sensor layer, acquiring data from an environment.  This layer detects, collects and processes information before transmitting it to the network layer.  
    \item \textit{Network layer} – performs IoT node collaborations in local and short range networks.  Handles data routing and transmission to different IoT hubs and devices over the internet.  Clouds, gateways, switches and routers use wireless protocols here.  
    \item \textit{Application layer} – should guarantee authenticity, integrity and confidentiality of data.  
\end{enumerate}

This three layer model is a worthwhile contribution though is perhaps a little basic given the complexity of IoT.  CISCO's IoT reference model \cite{ciscoiot} on the other hand contains seven layers, per Figure \ref{fig:ciscoiotmodel}.  CISCO contend IoT networks need their own communication and processing model, and \cite{ciscoiot} aims provide security for all processes at each level, plus secure movement and communications between those levels.  The CISCO reference model has been used in previous research \cite{7562568}, and is described as follows:

\begin{itemize}
    \item \textit{Layer 7 Collaboration and Processes} - people and business processes, e.g. identity management software.
    \item \textit{Layer 6 Application} – for reporting, analytics and control e.g. authentication/authorization software.
    \item \textit{Layer 5 Data Abstraction} – aggregation, access, and secure storage e.g. hardware and software solutions.
    \item \textit{Layer 4 Data Accumulation} – also involves storage e.g. tamper resistant software.
    \item \textit{Layer 3 Edge (or fog) Computing} - data element analysis and transformation where the network meets the cloud e.g. secure communications via protocols and encryption.
    \item \textit{Layer 2 Connectivity} – communication and processing units e.g. secure network access via hardware and protocols.
    \item \textit{Layer 1 Physical Devices and Controllers} – the IoT Edge devices and objects themselves.
\end{itemize}

\section{IoT topologies}

\cite{zegzhda} proposes three topologies:  1) point to point, 2) star and 3) mesh.  The literature indicates a preference for a mesh topology in IoT, with its decentralization using a gateway node, simple sensor nodes, and nodes that can work as both sensors and routers.  This decentralised mesh topology is recommended due to centralized setups having a weakness in providing a single point of failure if using a front-end proxy solution \cite{zegzhda, roman}.  Decentralising NoT is therefore attractive, where all nodes either participate in decision making and internal protocol, known as a flat configuration, or are divided into clusters of nodes, each with a cluster head to make decisions, known as a hierarchical configuration \cite{roman}.   Even using a central security manager represents a single point of failure and fixes network roles statically, in contrast to decentralized and distributed architecture being more dynamic \cite{rfc8576}.  The fundamental challenge of routing in IoT mesh networks comes from the requirement of maintaining routing information for each host in a multilink environment \cite{ndn}.  This is not an issue in traditional IP networks where routers or self-learning bridges can be deployed to provide infrastructural support for routing and forwarding.  However, in constrained IoT environments, the per-host routes are either maintained by every node in the mesh using routing protocols, which consumes lots of memory, or carried with the IP packet as source routes during forwarding which conflicts with the small MTU restriction.

\section{IoT protocols}
Convergence toward an IP-based communication stack is necessary as IoT has diverse wireless communications with gateway devices needed for protocol translation \cite{Elbouanani}.  Previous work discusses numerous protocols such as 6LoWPAN, IPv6 Routing Protocol for Low-Power and Lossy Networks (RPL), Datagram Transport Layer Security (DTLS), Constrained Application Protocol (CoAP), IPSec and Radio Frequency Identification (RFID).  Standardisation across these security protocols will be indispensable for the success of IoT \cite{keoh}.  

Firstly, consider the large number of IP addresses required in the IoT. IPv6 will ultimately be used instead of IPv4 as a result.  However with device limitations additional protocols are needed.  The Internet Engineering TaskForce (IETF) have developed 6LoWPAN and RPL\cite{Elbouanani}.  6LoWPAN uses compression to allow IPv6 packets to be sent over wireless networks made up of resource constrained devices.  6LoWPAN defines header compression to allow IPv6 packets to be sent between resource constrained devices \cite{Elbouanani}.  RPL has three modes of varying levels of security: unsecured, preinstalled and authenticated \cite{Elbouanani}.

CoAP is an application layer protocol to let devices communicate, using User Datagram Protocol (UDP).  It can be translated to HTTP for use over the web. DTLS is the basic building block for protecting CoAP \cite{rfc8576}.  DTLS however was designed for traditional computer networks rather than IoT, and DTLS imposes high overheads on IoT devices \cite{ndn}.  DTLS has been chosen as the channel security under CoAP for IoT \cite{keoh}.  DTLS was not designed for constrained environments though and is for point-to-point rather than publish-subscribe i.e. one-to-many.  It still has its weaknesses, with packet fragmentation and having to retransmit all messages in flight if one fails being drawbacks \cite{rfc8576}.  The optimisation of DTLS for CoAP over IoT is needed and IETF are working on a standard way of granting permissions and authorizing IoT to accept each other’s resources \cite{keoh}.  Eclipse's tinydtls library \cite{eclipsetiny} is the first open-source implementation of the protocol for IoT devices.  \cite{ndn} warns DTLS applies in IP-based apps but as a secure channel solution it does not fit into IoT for several reasons.  Firstly, there is the overhead of establishing a channel, and secondly both ends of the channel having to maintain the state of the channel until it closes.  This hinders memory usage when a device needs to communicate with many peers simultaneously in a densely-meshed network. 

\cite{keoh, rfc8576} suggest two other protocols for security aside from DTLS – 1) IPSec, for channel security via authentication headers and data security via encapsulating security payload traffic, and 2) IKEv2, which is used to establish IPSec.  IPSec is not supported for network layer security when using the 6LowPAN spec \cite{roman}, indicative of the standardisation issues.  Fragmentation due to a smaller MTU is problematic and open to attack \cite{rfc8576, ndn, keoh}.  At most 102 bytes are available for an IP packet after taking into consideration MAC frame header size and security.  Of this 102, another 48 bytes are needed for IPv6 and UDP headers, leaving just 64 bytes for application data and its security protection.  Hence fragmentation is needed though the trade-off is its vulnerability.

RFID is common with IoT, using tags containing antennae which are attached to devices in order to track and identify them via wireless/radio technology.  However, it is vulnerable, as outlined in the next section, and at time of writing there is a research challenge in connecting RFID devices over HTTPS.  MQTT (formerly MQ Telemetry Transport), created by OASIS, is a simple and lightweight protocol again suiting resource constrained devices \cite{Elbouanani}. MQTT, along with CoAP, is most commonly used in IoT \cite{andrea}. 

A larger critical mass of devices may be needed to reach a tipping point to achieve interoperable, standardised IoT, though it has been said designing a totally new protocol may seem like reinventing the wheel \cite{keoh}.  Moreover, given the billions of new IoT devices coming online, authentication based on digital certificates is very unlikely to scale to this level \cite{7575852}. 

\section{IoT vulnerabilities and attacks}

\subsection{IoT-specific attacks}

The cybersecurity Confidentiality – Integrity – Availability (CIA) triad also applies to IoT.  OWASP have a useful list of IoT attack surfaces which should be understood by manufacturers, developers, researchers and companies looking to deploy IoT in their organisations \cite{owasp17}.  Common IoT attacks are the sinkhole attack, the Sybil attack, the wormhole attack, the HELLO flood and acknowledgement spoofing \cite{1203362, wood, andrea}:

\begin{enumerate}
    \item \textit{Sinkhole attack} – all traffic is lured from an area through a compromised node, where selective forwarding can follow with the attacker deciding what data to allow through. 
    \item \textit{Sybil attack} – a single node presents multiple identities to others in the network, so an attacker can be in more than one place at once.
    \item \textit{Wormhole attack} – an attacker tunnels messages received in one part of the network over a low latency link and replays them in a different part.  
    \item \textit{HELLO flood} – here the attacker causes every node to mark it as their parent.  Most nodes will be out of range and this causes a lot of packets to be lost.  Routing loops can be set up via spoofing routing updates, with two nodes being attacked and redirecting packets to each other.
    \item \textit{Acknowledgement spoofing} - used for a selective forwarding attack, where an attacker strengthens/weakens networks links so packets are lost from a node.
\end{enumerate}

\subsection{Conventional network attacks}

Conventional network attacks can be executed at Layer 2 and Layer 1 of the CISCO seven-layer IoT model \cite{7562568}. Layer 2 connectivity attacks are:

\begin{enumerate}
    \item \textit{Eavesdropping / sniffing} – gains usernames, passwords, node identifiers and other useful data.
    \item \textit{Denial of Service (DoS) attacks} – jamming radio signals, using malicious nodes to refuse to route messages or redirect them to unwanted locations.
    \item \textit{Injecting fraudulent packets} – done via insertion (where malicious packets that seem legit are generated and sent), manipulation (when packets are captured then modified) and replication (where the attacker captures packets between two things to replay them).
    \item \textit{Routing attacks} – an attacker can spoof, redirect, misdirect or drop packets, for example with the aforementioned wormhole, HELLO flood and Sybil attacks.  
\end{enumerate}

Note that man-in-the-middle attacks are also possible in IoT if keying material is exchanged in the clear, or if device authentication is non-trivial and needs human interaction.  Layer 1 physical device attacks are:

\begin{enumerate}
   \item \textit{DoS attacks} – such as battery draining by an attacker sending lots of packets, outage attacks, or some other function to cause an edge device to stop performing normally.
   \item \textit{Node replication attacks} – the attacker adds a new node to an existing set by replicating another node’s ID.  Can lead to reduction in network performance and the attacker can further corrupt or misdirect packets that arrive at the replicated node. 
   \item \textit{Camouflage} - the attacker inserts a counterfeit edge node or attacks an authorized node to hide at the edge level.  Then it can obtain and manipulate packets, or passively analyse traffic.  
\end{enumerate}

\subsection{RFID attacks}

RFID is a weak protocol that allows person and device tracking \cite{vashi}, with IoT devices using RFID being too small to use complex security algorithms.  Cryptography can help, as can digital forgetting to protect privacy. The threats to RFID are significant \cite{xingmei, 7562568, rfc8576}:

\begin{enumerate}
    \item \textit{Replication attack} – copy or forge identical RFID labels.
    \item \textit{Channel Blocking attacks} – channel is occupied for a long time and legit communications can’t be transferred.
    \item \textit{Forgery attack} - a legit RFID label is obtained by using special hardware facilities of counterfeit.
    \item \textit{Impersonation attack} – attacker fakes a legit reader to steal or change RFID tag info.
    \item \textit{Tampering attack} – attacker will modify the info and pass it on to receiver.
    \item \textit{Tracking} – in close proximity a reader can read a tag.  Dangerous when combined with personal info.
    \item \textit{Inventorying} – info can be deduced from device tags
    \item \textit{DoS} – RF channels are jammed so the tags cannot be read by tag readers and the intended services become unavailable, e.g. locking down a whole building.
    \item \textit{Eavesdropping} – messages are intercepted/read/saved for future.
\end{enumerate}

\section{IoT mitigations and countermeasures}
\label{sec:mitigations}

By convention, a layered approach to cybersecurity is ideal and this again applies in IoT. IoT must be secured from hardware of nodes right through to applications \cite{roman}. One option is a holistic view to design security in at the OS level, using the devices hardware capabilities and extending up the device stack \cite{TANKARD201511}. Adding security to legacy devices, rather than solely focusing on devices to come, is important. Securing premises is obvious \cite{andrea}, though it is naive to stop there.  The computational capabilities of embedded systems in IoT should improve, and so some mitigations may be temporary \cite{keoh, rfc8576}, but this does not justify ignoring them.

\subsection{Choice of protocol}
Specifically proposed in \cite{rfc8576} as a countermeasure to DoS are DTLS and IPSec/IKEv2.  They implement return route checks based on cookies to delay state establishment until the initiating host address is verified.  DTLS together with IPSec/IKEv2 provide end-to-end security services including peer entity authentication, end-to-end encryption and integrity protection. \cite{wood} proposes Secure Implicit Geographic Forwarding (SIGF), a configurable secure routing protocol for NoT.  It does not use routing tables, preventing state corruption, wormholes and HELLO floods.  Still, this protocol is open to DoS and Sybil attacks.  To reduce packet loss, \cite{ndn} suggests legacy protocols should be redesigned to minimise use of IP multicast before they can be applied to IoT.  They say is it better for nodes to pull packets on-demand from a store where packets are buffered.  In addition, \cite{7562568} offers depattern-ing as a mitigation, where fake packets are inserted to fool the attacker.  \cite{rfc8576} also suggests puzzle-based approaches that forces the initiator to solve cryptographic puzzles of varying difficulty.  This should be used with care, as under attack conditions that reduce performance, devices may not be able to solve these puzzles and suffer exclusion. Secure routing is vital to acceptance and use of IoT, though current protocols are insecure with a new standard being needed \cite{7575852}, and we should be aware of attacks coming from more powerful devices outside the network that can break cryptography. 

\subsection{Application data security}
The IETF has suggested object-based security which secures the application data directly rather than securing the channel through which the data is transmitted \cite{7575852}.  Each object should have digital signatures so anyone receiving it can verify its validity.

\subsection{Intrusion detection systems}
\cite{raza} proposed an IoT specific IDS called SVELTE.  \cite{7562568} references SVELTE as one of the first IDS for IoT, with \cite{gendreau} seeming to go so far as saying that the prevention of unauthorised access to IoT will depend on intrusion detection capability of embedded devices.  This may be true though future protocol standards and legal regulations are still needed in parallel.

\subsection{RFID specific mitigations}
For attacks on RFID tags, \cite{7562568} suggests personal firewalls to examine all readers’ requests to read tags, along with the use of cryptography, though full encryption is difficult with IoT resource constraints.   Hash-based schemes are more widely used, where an RFID reader gets a hashed key from a locked RFID tag which it sends to a database.  The database returns a key to the reader which it uses to unlock the tag.

\subsection{Reducing risk with legislation}
\cite{verizon} view changing the law as a form of mitigation, discussing data protection laws and the trade-off between security and convenience.  Without legislation, manufacturers cannot be compelled to include security \cite{TANKARD201511}, and IoT will have to comply with regulatory industry frameworks at some point.  For example, it is expected 80\% of households will have energy meters by 2020 \cite{boban}.

\subsection{Choice of topology}
Per Section III, a decentralised topology is a logical mitigation, such as WirelessHART \cite{zegzhda}.

\section{Conclusions and future directions}
IoT security suffers from device constraints and a lack of standards, including IoT-specific protocols and channel-based security solutions.  Though non-trivial, research opportunities exist in the use of IDS in IoT, energy-saving cryptography techniques that are lightweight enough, and more broadly how to do more with less given such constrained resources.  Processing power doubles every two years according to Moore’s Law, so some mitigations that exist now could just be a stopgap until device resources catch up, although one could question if Moore's Law can really hold in IoT.  In addition, little work has covered the issue of device mobility in IoT - the rise in popularity of wearables, for example, poses complex issues of handling devices regularly leaving one NoT and joining another, not least for correlating and fingerprinting.  With attackers already exploiting the early adoption in a myriad of ways, IoT will continue to be a major attack surface given its sheer size coupled the resource challenges. The race already underway to protect legacy, existing and future devices through technology and robust legislation.

\ifCLASSOPTIONcaptionsoff
  \newpage
\fi



%\bibliographystyle{IEEEtran}
%\begin{thebibliography}{1}

 
\bibliographystyle{IEEEtran}
\bibliography{ccs-sample}

\end{document}